\documentclass[sigconf]{acmart}

\usepackage{pifont}
\usepackage[most]{tcolorbox}
\usepackage{booktabs}
\usepackage{tikz}
\usepackage{algorithm}
\usepackage{algpseudocode}
\usepackage{enumitem}
\usepackage{multirow}

\newcommand{\circmark}[1]{%
  \tikz[baseline=(char.base)]{%
    \node[shape=circle, fill=purple!60, inner sep=1.2pt,
          text=white, font=\small\bfseries] (char) {#1};}}
\newcommand{\tool}{\textsc{AgentSZZ}}

\AtBeginDocument{%
  }

\setcopyright{acmlicensed}
\copyrightyear{2026}
\acmYear{2026}
\acmDOI{XXXXXXX.XXXXXXX}
\acmConference[Preprint]{}{Under Submission}{Arxiv}
\acmBooktitle{}
\acmISBN{}
\acmISBN{978-1-4503-XXXX-X/2018/06}

\begin{document}
\title{\tool{}: Teaching the LLM Agent to Play Detective with Bug-Inducing Commits}

\author{Yunbo Lyu\textsuperscript{1}, Jieke Shi\textsuperscript{1}, Hong Jin Kang\textsuperscript{2}, Ratnadira Widyasari\textsuperscript{1}, Junda He\textsuperscript{1}, Yuqing Niu\textsuperscript{1},\\ Chengran Yang\textsuperscript{1*}, Junkai Chen\textsuperscript{1}, Zhou Yang\textsuperscript{3}, Julia Lawall\textsuperscript{4}, David Lo\textsuperscript{1}}

\affiliation{%
  \institution{%
    \textsuperscript{1}Singapore Management University, Singapore \quad
    \textsuperscript{2}University of Sydney, Sydney, Australia \quad
    \textsuperscript{*}Corresponding author \quad
    \textsuperscript{3}University of Alberta, Edmonton, Canada \quad
    \textsuperscript{4}Inria-Paris, Paris, France 
  }
  \country{}
}

\email{{yunbolyu, jiekeshi, ratnadiraw, yuqingniu, cryang, junkaichen, davidlo}@smu.edu.sg}
\email{hongjin.kang@sydney.edu.au, jundahe.2022@phdcs.smu.edu.sg, zy25@ualberta.ca, julia.lawall@inria.fr}
\renewcommand{\shortauthors}{Lyu et al.}

\begin{abstract}
The SZZ algorithm is the dominant technique for identifying bug-inducing commits and underpins many software engineering tasks, such as defect prediction and vulnerability analysis. 
Despite numerous variants, including recent LLM-based approaches, performance remains limited on developer-annotated datasets (e.g., recall of 0.552 on the Linux kernel). 
A key limitation is the reliance on \textit{git blame}, which traces line-level changes within the same file, failing in common scenarios such as ghost and cross-file cases—making nearly one-quarter of bug-inducing commits inherently untraceable. 
Moreover, current approaches follow fixed pipelines that restrict iterative reasoning and exploration, unlike developers who investigate bugs through an interactive, multi-tool process.

To address these challenges, we propose \tool{}, an agent-based framework that leverages LLM-driven agents to explore repositories and identify bug-inducing commits. 
Unlike prior methods, \tool{} integrates task-specific tools, domain knowledge, and a ReAct-style loop to enable adaptive and causal tracing of bugs. 
A structured compression module further improves efficiency by reducing redundant context while preserving key evidence.
Extensive experiments on three widely used datasets show that \tool{} consistently outperforms state-of-the-art SZZ algorithms across all settings, achieving F1-score gains of up to 27.2\% over prior LLM-based approaches. 
The improvements are especially pronounced in challenging scenarios such as cross-file and ghost commits, with recall gains of up to 300\% and 60\%, respectively. 
Ablation studies show that task-specific tools and domain knowledge are critical, while compression tool outputs reduces token consumption by over 30\% with negligible impact.
Replication package is available.

\end{abstract}



\maketitle

\section{Introduction}
\label{sec:intro}

SZZ algorithms have been proposed since 2005~\cite{jacek2005when} and have become a foundational technique with broad impact in the software engineering (SE) community~\cite{sigsoft-impact-paper-award}. 
Given a bug-fixing commit, SZZ identifies the bug-inducing commits that introduced the defect. 
They have been widely used in many downstream tasks such as defect prediction~\cite{fan2019impact,hata2012bug,tan2015online,yan2022just,pascarella2019fine}, analysis of how bugs are introduced~\cite{tufano2017empirical,bavota2015four,chen2019extracting}, and vulnerability version identification~\cite{bao2022v,chen2025vulnerability}. 
More recently, security code generation benchmarks also rely on SZZ algorithms to identify vulnerability-inducing commits~\cite{chen2025secureagentbench}.

Many variants of SZZ algorithms have been proposed to improve the accuracy of bug-inducing commit identification. 
AG-SZZ~\cite{kim2006automatic} filters cosmetic changes such as blank lines and comments; MA-SZZ~\cite{da2016framework} excludes meta, branch, and property changes; and RA-SZZ~\cite{neto2018impact} detects and filters refactoring changes. 
More recently, LLM4SZZ~\cite{tang2025llm4szz} leverages LLMs to assess blame candidates with expanded context, achieving state-of-the-art (SOTA) results across multiple datasets. 
Despite these advances, performance remains far from satisfactory on developer-annotated datasets~\cite{lyu2024evaluating,rosa2021evaluating,wen2019exploring}. 
For example, LLM4SZZ achieves a recall of only 0.552 on the Linux kernel dataset, indicating that nearly half of the bug-inducing commits remain unidentified.

The \textbf{fundamental limitation} is that existing SZZ algorithms rely heavily on \textit{git blame} to trace modified lines in bug-fixing commits. 
As a result, they fail to identify bug-fixing commits without deleted or modified lines (i.e., ghost commits~\cite{rezk2021ghost}) and bug-inducing commits located in different files (i.e., cross-file cases~\cite{lyu2024evaluating}). 
Prior studies show that, in the Linux kernel dataset, 17.47\% of bug-fixing commits are ghost commits and 7.20\% are cross-file cases~\cite{lyu2024evaluating}, meaning nearly one-quarter of bug-inducing commits are inherently untraceable by blame-based methods.
However, in practice, developers are not limited to a single tool, but \textbf{use diverse tools} (e.g., \textit{git log}) to gather context and understand bugs~\cite{parnin2011automated,latoza2006maintaining}. 
Current SOTA LLM4SZZ~\cite{tang2025llm4szz} still follow fixed pipelines that predefine what to inspect, limiting iterative reasoning and exploration. 
In contrast, developers follow an \textbf{information foraging process}~\cite{lawrance2013how}, exploring promising paths and backtracking when encountering uninformative changes.

Agent-based approaches are a natural fit for this problem, as they can simulate developers' dynamic and interactive investigation processes, enabling models to actively explore repositories and follow information scents~\cite{zhang2023repocoder,yang2024swe,zhang2024autocoderover}. 
However, \textbf{they are not a silver bullet}. 
Our exploration with mini-SWE-Agent~\cite{yang2024swe}, a powerful SWE agent with flexible bash interaction, shows poor performance. 
Several challenges remain: 
\ding{182} agents struggle with tool selection, 
\ding{183} actions lack domain-specific guidance, and 
\ding{184} tool outputs can be large, leading to excessive token consumption, increased latency, and noisy context that may hinder reasoning.

\begin{figure*}[t]
  \centering
  \includegraphics[width=0.95\linewidth]{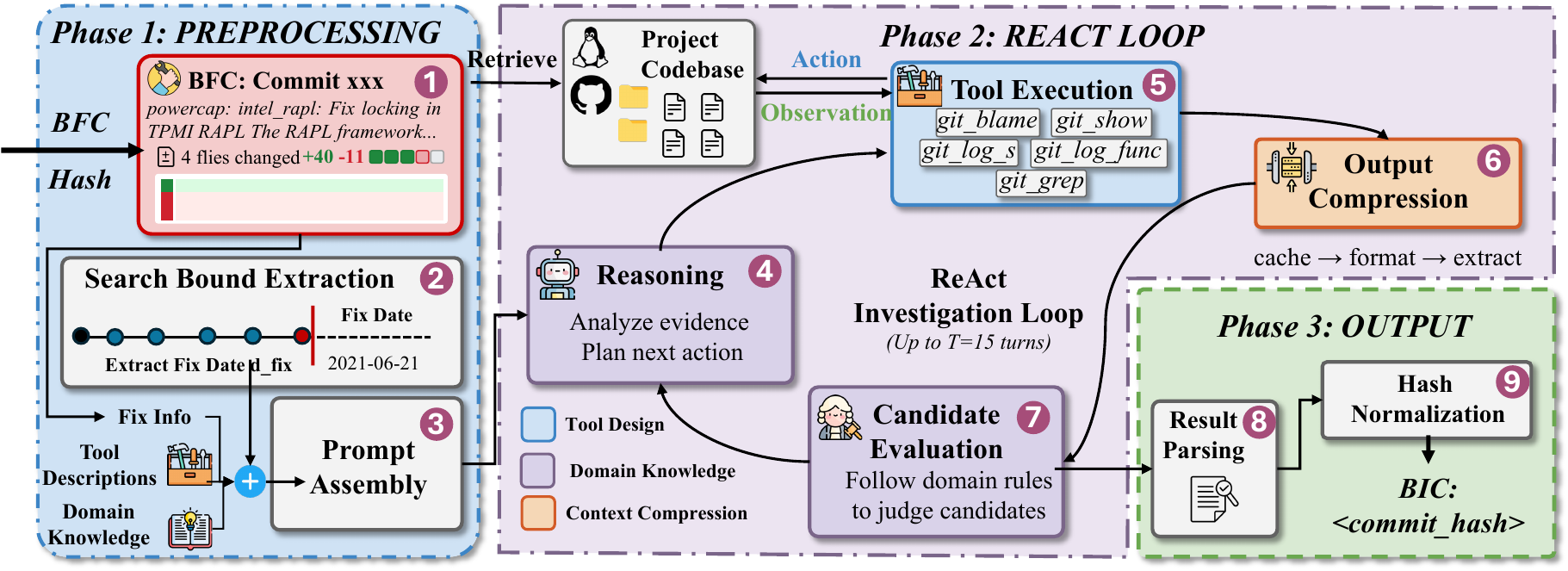}
  \caption{Overview of \tool{}, illustrating the end-to-end workflow from preprocessing (1-3), through the ReAct investigation loop (4-7), to output generation (8-9). 
  Given a bug-fixing commit (BFC), the agent investigates the repository through tool interactions guided by domain knowledge, with context compression, and outputs the identified bug-inducing commit (BIC).
  }
  \label{fig:overview}
\end{figure*}

To address these challenges, we propose \tool{}, as shown in Figure~\ref{fig:overview}, an agent-based framework that uses LLM-driven agents to interact with repositories and identify bug-inducing commits akin to a detective. 
Unlike prior approaches that rely primarily on \textit{git blame}, \tool{} interacts with repositories through five task-specific tools designed for bug-inducing commit identification. 
Given a fix commit, it iteratively explores the repository via a ReAct-style loop~\cite{yao2022react}, using GPT-5-mini as the backbone model with a maximum of 15 interaction turns, leveraging these tools and domain knowledge to trace causal evidence. 
A structured compression module further improves efficiency by reducing redundant and noisy context while preserving key information.

The evaluation results on three widely used datasets show that \tool{} consistently outperforms SOTA SZZ algorithms across all settings. 
\tool{} achieves substantial gains, improving F1-score by up to 27.2\% over prior SOTA LLM4SZZ, and maintains strong performance even under the same backbone model, indicating that the improvements stem from the agent design rather than the choice of LLM. 
These gains are especially pronounced in challenging scenarios such as cross-file and ghost commits, where recall improves by up to 300\% and 60\%, respectively, highlighting the ability of \tool{} to improve the fundamental limitations of blame-based methods. 
Ablation studies further reveal that task-specific tools and domain knowledge are central to performance, while the compression module significantly improves efficiency by reducing token consumption by over 30\% with negligible impact. 
\tool{} also generalizes well across datasets and programming languages without dataset-specific tuning.

In summary, we make the following contributions: 
(1) We propose \tool{}, an agent-based framework that leverages LLM-based agents to interact with repositories and identify bug-inducing commits, enabling \textbf{adaptive and causal tracing} beyond the limitations of existing SZZ algorithms. 
(2) We design five task-specific tools, encode domain knowledge, and introduce a structured compression module to improve the agent's reasoning capability and efficiency in repository exploration. 
(3) We conduct extensive experiments on three widely used datasets, showing that \tool{} significantly outperforms SOTA SZZ algorithms, particularly in challenging scenarios such as ghost commits and cross-file cases.
(4) We provide a replication package including code and data.

The remainder of this paper is organized as follows. 
Section~\ref{sec:background} provides background on SZZ algorithms and LLM-based agents for software engineering. 
Section~\ref{sec:methodology} presents the design of \tool{}, Section~\ref{sec:results} describes the experimental setup, and Section~\ref{sec:results} reports the evaluation results. 
Section~\ref{sec:discussion} offers further analysis and discussion, and Section~\ref{sec:conclusion} concludes the paper.



\section{Background and Related Work}
\label{sec:background}

\subsection{SZZ Algorithms}

The SZZ algorithm~\cite{jacek2005when} is the foundational approach for automated BIC identification. 
Given a bug-fixing commit, it traces deleted or modified lines back to the commits that last touched them, treating those commits as BIC candidates. 
The original implementation operates on CVS repositories using the annotate command; modern implementations adapt this principle to Git via \textit{git blame}. 
While conceptually simple, the algorithm suffers from false positives, as not all changes to the fixed lines are semantically meaningful.

A series of refinements have been proposed to address these limitations. 
AG-SZZ~\cite{kim2006automatic} filters cosmetic changes such as blank lines and comments, and introduces annotation graphs to track line origins more precisely across revisions. 
MA-SZZ~\cite{da2016framework} extends this by excluding meta-changes (e.g., branch merges and property changes) that do not alter program behavior. 
RA-SZZ~\cite{neto2018impact} further detects and filters refactoring operations using RefDiff~\cite{silva2017refdiff} and Refactoring Miner~\cite{tsantalis2018accurate}, though this capability is limited to Java. 
When multiple candidates remain, R-SZZ and L-SZZ~\cite{davies2014comparing} apply simple heuristics, selecting the most recent commit or the one with the most changed lines, respectively. 

More recent work shifts from heuristic filtering to learned or model-based ranking. 
Neural-SZZ~\cite{tang2023neural} embeds changed lines using CodeBERT~\cite{feng2020codebert} and ranks candidates via a heterogeneous graph attention network, capturing semantic relationships beyond rule-based methods. 
LLM4SZZ~\cite{tang2025llm4szz} further leverages LLMs to assess blame candidates with expanded context, achieving state-of-the-art results across multiple datasets. 
However, it still operates as a static pipeline without iterative investigation.

\subsection{LLM-Based Agents for Software Engineering}
\label{subsec:agent}

The application of LLMs in SE has evolved from static, single-turn code generation to dynamic, autonomous agents~\cite{liu2026llmagentsurvey,he2025llm}. 
Unlike traditional LLM applications that rely solely on internal model weights, LLM-based agents are augmented with reasoning frameworks (e.g., ReAct~\cite{yao2022react}), memory mechanisms, and the ability to interact with external environments via tools~\cite{xi2025rise}. 
This evolution enables agents to tackle complex SE tasks across the Software Development Life Cycle, shifting from snippet-level synthesis to repository-level reasoning and maintenance~\cite{liu2026llmagentsurvey}.

Recent work on agents designed to navigate large codebases and resolve real-world issues, as exemplified by SWE-bench~\cite{jimenez2023swe}.
Systems such as Devin~\cite{cognition2024devin} and OpenHands~\cite{wang2024openhands} demonstrate end-to-end patch generation, while SWE-agent~\cite{yang2024swe} introduces an Agent-Computer Interface (ACI) to bridge LLMs and terminal environments by structuring tool interactions. 
Lightweight variants such as mini-SWE-agent (a lightweight variant of SWE-agent)~\cite{yang2024swe} demonstrate that minimal bash-based agents can achieve competitive performance. 
Other approaches enhance repository interaction through structured search or execution mechanisms, including AST-based navigation (AutoCodeRover~\cite{zhang2024autocoderover}) and executable action spaces (CodeAct~\cite{wang2024executable}). 
In contrast, pipeline-based methods such as Agentless~\cite{xia2025demystifying} simplify execution by replacing autonomous planning with a fixed pipeline.
Despite these advances, existing agents either rely on general-purpose exploration (e.g., bash-based interaction) or predefined pipelines, lacking task-specific tool design and domain knowledge for precise bug-inducing commit identification. 
\tool{} addresses this gap by combining structured, task-specific tools with domain knowledge to enable accurate bug-inducing commit identification.

\section{\tool{} Approach}
\label{sec:methodology}

To address the limitations of existing SZZ algorithms and the challenges of applying agent-based approaches to bug-inducing commit identification, we propose \tool{}, an agent-based framework that leverages LLM-based agents to interact with repositories and identify bug-inducing commits, as illustrated in Figure~\ref{fig:overview}.
\tool{} performs \textbf{adaptive investigation} that more closely resembles how developers analyze bugs, in contrast to prior SZZ approaches that follow fixed pipelines. 
Inspired by the ReAct paradigm~\cite{yao2022react}, the agent dynamically selects investigation strategies based on intermediate findings, enabling flexible repository exploration. 
Instead of relying on syntactic line matching, it \textbf{reasons about causal relationships} between commits and the bug, distinguishing superficial changes (e.g., refactorings) from those that introduce faulty logic. 
This enables \tool{} to handle challenging scenarios such as cross-file cases and refactoring-obscured bugs.


\subsection{Design Process}
\label{subsec:design_process}

The components of \tool{} were derived through an iterative, case-driven design process rather than predefined. 
To guide this process, we randomly sampled 50 cases from the Linux kernel dataset~\cite{lyu2024evaluating}, excluding those used in the final evaluation~\cite{tang2025llm4szz}. 
We did not scale up this set due to the high cost of in-depth case analysis, as our goal was to perform detailed investigations to understand key challenges and inform component design. 
We conducted our study on the Linux kernel, a large and complex codebase that enables diverse and challenging cases. 
We investigated these cases while interacting with a LLM~\cite{anthropic2026claudeopus46} to simulate the agent's behavior.

Through this process, we iteratively refined the tool design, domain knowledge, and context management. 
Initially, we employed \texttt{git\_blame} and \texttt{git\_show} as core tools following standard SZZ practices. 
These tools were implemented as thin wrappers with only essential parameters (e.g., commit hash and file path), which often produced excessive outputs and led to inefficient investigation. 
To address this, we introduced scoped parameters (e.g., line ranges, file filters, and search bounds) to constrain the search space. 
We further expanded the tool set to handle more complex scenarios by adding \texttt{git\_log\_s} for locating commits that introduce or remove specific identifiers, \texttt{git\_grep} for repository-wide search, and \texttt{git\_log\_func} for function-level history tracing. 
We exclude \texttt{git bisect}, as it requires compiling and executing.

Beyond tool refinement, we distilled domain knowledge from these investigations and existing literature. 
This includes heuristics such as starting with blame as a default entry point while allowing the agent to switch to broader search tools (e.g., \texttt{git\_log\_s}) when necessary. 
We also encourage continued blame tracing when encountering refactoring~\cite{dacosta2017framework} or meta-changes~\cite{da2016framework} (e.g., merges), preventing premature convergence to non-causal commits. 
These heuristics guide deeper exploration and enable the identification of bug-inducing commits that require non-trivial investigation.
We further observed that tool outputs can be large, introducing unnecessary noise and increasing token consumption. 
This observation motivated the design of structured context compression to preserve essential evidence while reducing token usage.

\subsection{Approach Overview}
\label{subsec:overview}

As shown in Figure~\ref{fig:overview}, the overall workflow of \tool{} into three main phases: \textit{Preprocessing}, \textit{ReAct investigation loop}, and \textit{Output}.

\textbf{Preprocessing.} 
Given a fix commit, \tool{} retrieves the commit message and diff of the BFC \circmark{1}. 
Irrelevant information (e.g., \texttt{Reviewed-by} tags) is filtered to produce a concise content for the agent. 
To prevent information leakage, we remove any content that may reveal the BIC, including the \texttt{Fixes} tag in the Linux dataset~\cite{lyu2024evaluating} and BIC hashes in GitHub commit titles~\cite{rosa2021evaluating}. 
We then extract the fix commit date and use it as a hard temporal upper bound for all subsequent search operations \circmark{2}. 
Finally, fix information, tool descriptions, domain knowledge, and search constraints are assembled into the initial context for the LLM agent \circmark{3}.

\textbf{ReAct Investigation Loop.}
The agent then enters a multi-turn investigation loop following the ReAct paradigm~\cite{yao2022react}. 
This iterative observe--reason--act cycle enables adaptive exploration of the repository, with observations and intermediate reasoning accumulated across turns.
We set a maximum of 15 turns, following prior agent-based approaches~\cite{xie2024osworld,liu2025process,team2026kimi}. 
Given the initial context, the agent first analyzes the available information and plans the next action based on domain knowledge \circmark{4}. 
It then invokes one of the \textbf{task-specific tools} to interact with the repository and gather observations \circmark{5}. 
The tool output is subsequently processed by the context compression module to reduce token consumption \circmark{6}. 
Based on the compressed observations, the agent evaluates candidate commits according to domain knowledge and updates its belief about the BIC \circmark{7}. 
This process repeats until the agent identifies a BIC candidate with sufficient evidence or reaches the turn limit, allowing it to iteratively refine its search and adapt its strategy when initial approaches prove unfruitful.

\textbf{Output.}
Once the agent identifies a BIC candidate or reaches the turn limit, it produces a structured output containing the predicted BIC hash, confidence level, and supporting reasoning. 
\tool{} then parses this output \circmark{8} and normalizes the commit hash \circmark{9}. 
The extracted hash is sanitized to remove non-hexadecimal characters and verified against the repository using \texttt{git rev-parse}. 
If the full hash cannot be resolved (e.g., due to hallucination), 
If no valid commit can be resolved, the prediction is discarded.

\subsection{Task-Specific Tools}
\label{subsec:tools}

Table~\ref{tab:tools} summarizes the five tools provided to the agent. Instead of exposing raw git commands, we design a task-specific interface that constrains the agent's action space to operations relevant to bug-inducing commit identification.

\begin{table}[t]
\caption{Task-specific tools provided to the \tool{} agent.
         Parameters marked with * are required.
         These tools are designed to cover the key operations involved in tracing bug-inducing commits.}
\label{tab:tools}
\centering
\small
\begin{tabular}{@{}ll@{}}
\toprule
\textbf{Tool} & \textbf{Parameters} \\
\midrule
\texttt{git\_show}      & \texttt{commit*}, \texttt{file\_filter}, \texttt{stat\_only}, \texttt{context\_lines} \\
\texttt{git\_blame}     & \texttt{file\_path*}, \texttt{commit}, \texttt{line\_start}, \texttt{line\_end} \\
\texttt{git\_log\_s}    & \texttt{search\_string*}, \texttt{path}, \texttt{after}, \texttt{before} \\
\texttt{git\_log\_func} & \texttt{function\_name*}, \texttt{file\_path*}, \texttt{after}, \texttt{before} \\
\texttt{git\_grep}      & \texttt{search\_string*}, \texttt{commit}, \texttt{path} \\
\bottomrule
\end{tabular}
\end{table}

\textbf{Tool Abstraction.} 
An unrestricted git interface exposes hundreds of subcommands and flag combinations, many of which are irrelevant to bug-inducing commit identification. 
We distill this space into five tools, each supporting a distinct operation in the investigation process (Section~\ref{subsec:design_process}). 
\texttt{git\_blame} traces line-level ownership to identify the commit that last modified a given line. 
\texttt{git\_show} retrieves the commit message and diff for semantic classification (e.g., distinguishing refactoring from bug inducing). 
\texttt{git\_log\_s} (pickaxe search) and \texttt{git\_log\_func} (function-level history) support tracing code origins across file renames and structural changes. 
\texttt{git\_grep} searches the codebase at a specific revision to locate definitions, call sites, and cross-file references. 
All five tools wrap standard Git commands (\texttt{git blame},
\texttt{git show}, \texttt{git log -S}, \texttt{git log -L}, and
\texttt{git grep}), but are exposed as function-calling schemas via the
API \texttt{tools} parameter, restricting the agent to invoking only
these predefined operations with schema-conforming arguments.

\textbf{Scoped Parameters.} 
Each tool exposes a focused set of parameters that constrain its output scope, enabling more targeted queries. 
\texttt{git\_blame} supports \texttt{line\_start} and \texttt{line\_end} to focus on modified lines, while \texttt{git\_show} provides a \texttt{file\_filter} to restrict diff output to specific files. 
The search tools (\texttt{git\_log\_s}, \texttt{git\_log\_func}) support temporal scoping via date parameters, allowing exploration within relevant history. 
\texttt{git\_grep} accepts a \texttt{commit} parameter to search the codebase at a specific revision, and a \texttt{path} filter to narrow results to relevant directories. 
These parameters remain optional, allowing the agent to apply them based on context while preserving flexibility within a bounded search space.




\subsection{Domain Knowledge}
\label{subsec:domain_knowledge}

\begin{figure}[t]
  \centering
  \includegraphics[width=0.95\linewidth]{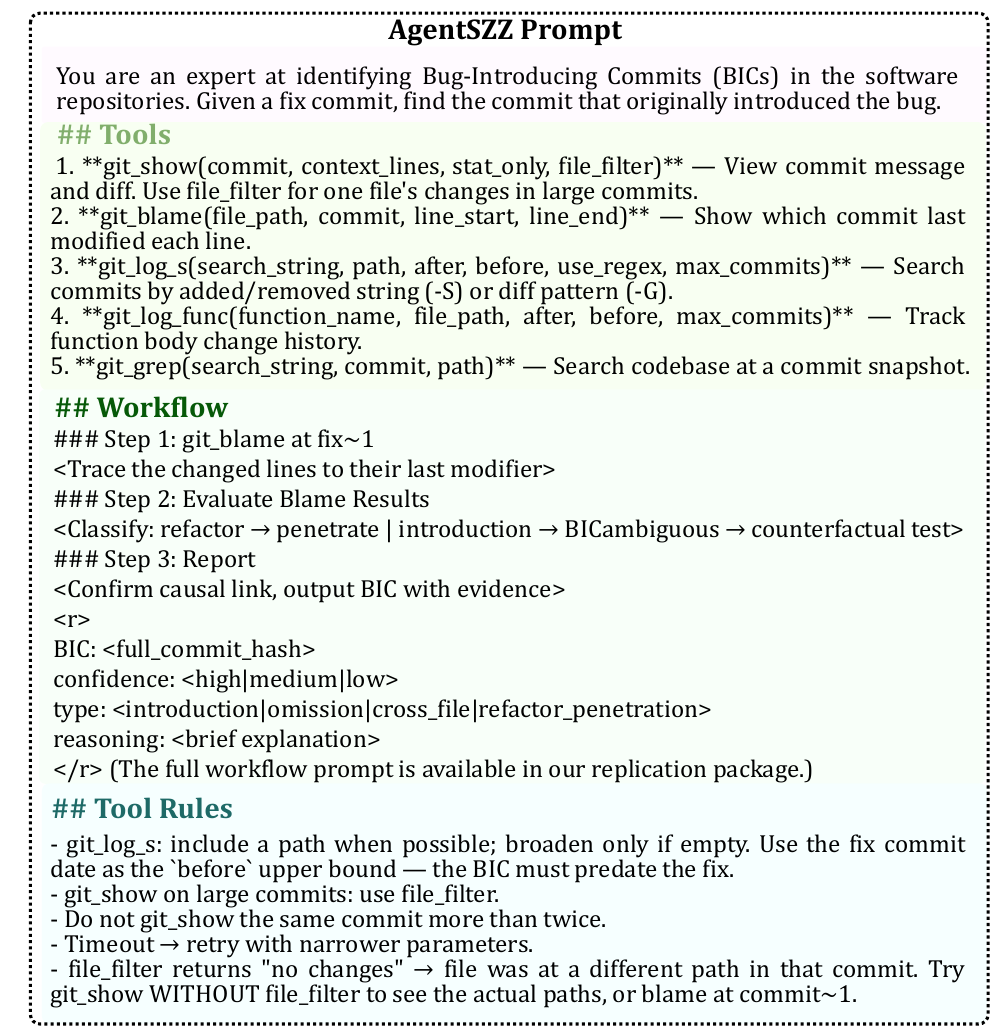}
  \caption{Prompt for \tool{}, encoding domain knowledge for Bug-Inducing Commit (BIC) investigation.}
  \label{fig:prompt}
\end{figure}

The agent's investigation is guided by a structured execution prompt that encodes domain knowledge distilled from our design process (Section~\ref{subsec:design_process}) and existing literature. 
The prompt of \tool{} (Fig.~\ref{fig:prompt}) consists of three components: a task description, tool specifications (Section~\ref{subsec:tools}), and domain knowledge that defines both high-level investigation guidance and operational rules. 

textbf{Investigation Workflow.}
The domain knowledge encodes a three-step investigation strategy as high-level guidance, while the agent retains full autonomy over tool selection, invocation order, and query parameters at each step.
\textbf{Step~1 (Entry point selection).}
The domain knowledge recommends starting with \texttt{git\_blame} on the modified or deleted lines at the parent of the bug-fixing commit (\texttt{fix\textasciitilde1}). 
For ghost commits (where the bug-fixing commit does not contain modified or deleted lines), our domain knowledge suggests blaming surrounding context lines. 
Moreover, we suggest that the agent use \texttt{git\_log\_s} or \texttt{git\_log\_func} to trace the history of relevant identifiers when blaming surrounding context does not work. 
For cross-file cases, we suggest that the agent may instead begin with broader search tools (e.g., \texttt{git\_log\_s}, \texttt{git\_grep}).
\textbf{Step~2 (Candidate analysis).}
Candidate commits surfaced during exploration are classified into two categories: 
(i)~\textit{non-semantic changes}, including refactoring and meta-changes (e.g., merge, branch), which are bypassed via iterative re-blame at the candidate's parent commit to reach the true origin, following established SZZ practices such as MA-SZZ~\cite{dacosta2017framework} and RA-SZZ~\cite{neto2018impact}; 
and (ii)~\textit{semantic introductions}, which introduce or modify logic and are treated as likely bug-inducing commits. 
When multiple semantic candidates exist, the prompt provides a counterfactual heuristic for disambiguation: ``would the bug exist if this commit had not been introduced?''
\textbf{Step~3 (Causal confirmation).}
Before reporting, the agent verifies a direct causal link between the selected candidate and the bug, ensuring it is causally responsible rather than merely correlated.

\textbf{Rules.}
The prompt further encodes lightweight operational constraints that can be applied at the agent's discretion. 
These include prioritizing path-restricted queries before broadening scope, avoiding redundant inspection (e.g., limiting repeated \texttt{git\_show} calls on the same commit), retrying with narrower parameters after timeouts, and accounting for file path changes across commits. 
It also encodes a key semantic distinction: commits that merely \textit{expose} or \textit{trigger} a pre-existing bug are not considered BICs; only commits that \textit{introduce} faulty logic qualify.

\subsection{Context Compression}
\label{subsec:compress}

To manage context consumption during multi-turn investigation, we design a multi-layer compression pipeline (Algorithm~\ref{alg:compression}) that reduces tool output size while preserving high-signal evidence. 

\begin{algorithm}[t]
\caption{Structured Context Compression}
\label{alg:compression}
\begin{algorithmic}[1]
\Require Tool name $t$, arguments $args$, fix date $d_{fix}$, 
         threshold $\tau$
\Ensure Compressed output $O'$
\State \textit{// Search bound enforcement}
\If{$t \in \{\text{log\_s}, \text{log\_func}\}$ \textbf{and} 
    $args.\text{before} > d_{fix}$}
    \State $args.\text{before} \gets d_{fix}$
\EndIf
\State \textit{// Layer 1: Cache deduplication}
\If{$\Call{Cache.Has}{t, args}$}
    \State $O_{\text{raw}} \gets \Call{Cache.Get}{t, args}$
\Else
    \State $O_{\text{raw}} \gets \Call{ExecuteTool}{t, args}$
    \If{$O_{\text{raw}}$ is timeout}
        \State Inject hint: ``\textit{Retry with narrower 
               parameters}''
        \State \Return timeout message
    \EndIf
    \State $\Call{Cache.Store}{t, args, O_{\text{raw}}}$
\EndIf
\State \textit{// Layer 2: Formatted output}
\State $O \gets \Call{Format}{t, O_{\text{raw}}}$:
\State \hspace{1em} Reformat raw output into concise 
       representation
\State \hspace{1em} Strip metadata trailers; truncate at 
       max line limit
\State \textit{// Layer 3: Structured extraction}
\If{$|O| \leq \tau$} \Return $O$
\EndIf
\State $O' \gets \Call{Extract}{t, O}$:
\State \hspace{1em} \textbf{show}: Strip diff context lines 
       (keep only \texttt{+}/\texttt{-}/\texttt{@@}); 
       head-tail truncate at $K_1$
\State \hspace{1em} \textbf{blame}: Deduplicate commits into 
       summary header; head-tail truncate at $K_2$
\State \hspace{1em} \textbf{log}: Keep first $K_3$ entries
\State \hspace{1em} \textbf{grep}: Group by file; keep first 
       $K_4$ matches
\State \Return $O'$
\end{algorithmic}
\end{algorithm}

\textbf{Search Bound Enforcement.} 
Before tool execution, the system automatically caps the \texttt{before} parameter of temporal search tools to the fix commit date $d_{fix}$. 
Since a bug-inducing commit must predate its fix, this removes irrelevant results without requiring the agent to explicitly enforce this constraint.

\textbf{Layer 1: Cache Deduplication.} 
A call-level cache keyed by tool name and arguments returns previously computed results when queries are repeated, preventing duplicate outputs from accumulating across turns.
\textbf{Timeout Recovery.} 
If a tool call times out, the system injects targeted hints (e.g., ``\textit{Retry with a smaller line range}'') to guide the agent toward narrower queries, avoiding repeated expensive operations.

\textbf{Layer 2: Formatted Outputs.} 
Each tool applies internal formatting before returning results to improve readability and reduce noise. 
\texttt{git\_blame} converts git's verbose porcelain format into a concise \texttt{L\{n\}: \{hash\} | \{code\}} representation with a commit legend. 
All commit-displaying tools filter out metadata trailers (e.g., \texttt{Signed-off-by}, \texttt{Reviewed-by}) that are irrelevant to bug-inducing commit identification. 
Additionally, each tool enforces a maximum output length (100--300 lines) with truncation notifications, enabling the agent to iteratively refine queries with narrower scopes.
The truncation thresholds (100--300 lines) are set based on the typical output volume of each tool: 
\texttt{git\_grep} produces many short matching lines and is capped at 100 to avoid flooding the context with repetitive results, while \texttt{git\_log\_func} includes inline diffs that require more space and is allowed up to 300 lines. 
When output exceeds the threshold, the tool appends a truncation notification, prompting the agent to retry with narrower parameters (e.g., adding a path filter or line range). We did not perform systematic tuning of these thresholds; they are designed to balance information completeness with context window efficiency.

\textbf{Layer 3: Structured Extraction.} 
When formatted outputs exceed a character threshold $\tau = 3000$, a deterministic, tool-specific extractor further compresses them. 
For \texttt{git\_show}, diff context lines are removed, retaining only change markers (\texttt{+}/\texttt{-}/\texttt{@@}) and file headers with head-tail truncation at $K_1$ lines. 
For \texttt{git\_blame}, repeated commits are deduplicated into a summary header, followed by head-tail truncation at $K_2$ lines. 
For \texttt{git\_log} variants, only the first $K_3$ entries are retained, and for \texttt{git\_grep}, matches are grouped by file before limiting to $K_4$ entries. 
These rules preserve structurally important information (e.g., commit hashes, line numbers, and changed code) while aggressively removing redundant context.

\subsection{Successful Example}
\label{subsec:successful_example}

\begin{figure}[]
  \centering
  \includegraphics[width=0.95\linewidth]{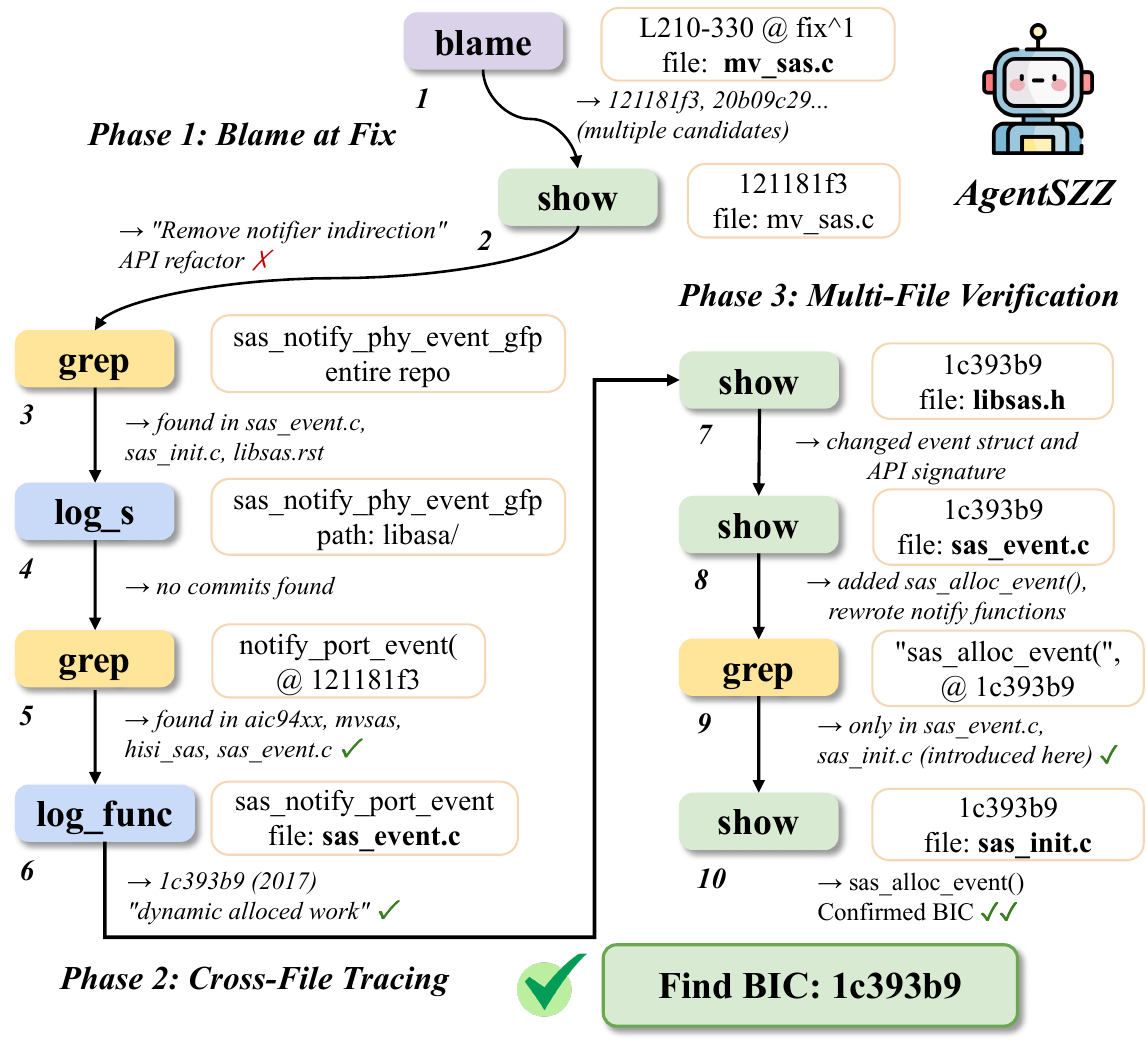}
  \caption{A successful case of \tool{} identifying a cross-file bug-inducing commit.}
  \label{fig:successful_example}
\end{figure}

We illustrate \tool{} through a cross-file bug in the Linux kernel's SCSI subsystem.\footnote{https://github.com/torvalds/linux/commit/feb18e900f0048001ff375dca639eaa327ab3} 
Although the fix modifies \texttt{mvsas/mv\_sas.c}, the root cause lies in the core library \texttt{libsas/sas\_event.c}, where the event handling mechanism was rewritten four years earlier. 
Fig.~\ref{fig:successful_example} shows the 10-turn investigation trajectory.

\textbf{Phase 1: Blame at fix (Turns 1--2).} 
Following the blame-first strategy, the agent blames the changed lines in \texttt{mv\_sas.c} at the parent commit and identifies commit \texttt{121181f3} as a frequent modifier. 
\texttt{git\_show} reveals this commit to be an API refactoring (``Remove notifier indirection'') that replaces indirect callbacks with direct calls—a mechanical change rather than the root cause.

\textbf{Phase 2: Cross-file tracing (Turns 3--6).} 
Recognizing that the bug lies beyond the driver file, the agent shifts to cross-file exploration. 
It uses \texttt{git\_grep} to locate sas\_notify\_phy\_event\_gfp across the repository, identifying references in \texttt{sas\_event.c}, \texttt{sas\_init.c}, and related files.
After an unsuccessful \texttt{git\_log\_s} search, the agent adapts by tracing the pre-refactoring function \texttt{notify\_port\_event}. 
Using \texttt{git\_log\_func}, it identifies commit \texttt{1c393b9} (2017) as the change that rewrote the event handling logic.

\textbf{Phase 3: Multi-file verification (Turns 7--10).} 
The agent verifies the candidate across multiple files. 
It confirms changes to the API and data structures in \texttt{libsas.h}, validates the shift to dynamic allocation in \texttt{sas\_event.c}, and uses \texttt{git\_grep} to confirm that \texttt{sas\_alloc\_event} was introduced in this commit. 
Finally, it inspects \texttt{sas\_init.c} to complete the verification.

The agent correctly identifies \texttt{1c393b9} as the bug-inducing commit with high confidence. 
This case highlights three key strengths of \tool{}: 
(i) \textit{domain knowledge} enables effective refactor penetration and strategy adaptation; 
(ii) \textit{task-specific tools} support cross-file tracing beyond line-based SZZ; 
and (iii) \textit{context compression} maintains tractable context across multi-turn investigation. 
This trajectory illustrates how \tool{} progressively shifts from local blame to broader repository exploration when local evidence becomes insufficient. 
Notably, the BIC resides in a different file and predates the fix by four years—conditions under which traditional SZZ approaches typically fail.

\section{Study Design}
\label{sec:design}

\subsection{Research Questions}
To evaluate \tool{}, we design the following research questions:

\begin{itemize}
    \item \textbf{RQ1.} How effective is \tool{} at identifying BICs compared to baselines?
    \item \textbf{RQ2.} How efficient is \tool{} in terms of time and cost?
    \item \textbf{RQ3.} What is the contribution of each key component of \tool{}?
    \item \textbf{RQ4.} Can \tool{} identify BICs in challenging scenarios?
\end{itemize}

In RQ1, we evaluate the effectiveness of \tool{} by comparing it with a range of existing SZZ algorithms on the task of identifying bug-inducing commits across multiple datasets. 
In RQ2, we assess the efficiency of \tool{} in terms of runtime, token consumption, and monetary cost, and compare it against the state-of-the-art LLM-based approach LLM4SZZ. 
In RQ3, we conduct an ablation study to quantify the contribution of each key component of \tool{} (i.e., task-specific tools, domain knowledge, and context compression) to its overall performance and efficiency. 
In RQ4, we further analyze the effectiveness of \tool{} in challenging scenarios, with a focus on cross-file and ghost commit cases that are known to be difficult for traditional SZZ methods~\cite{lyu2024evaluating,rezk2021ghost}.

\subsection{Evaluation Dataset}
\label{subsec:dataset}

To evaluate \tool{}, we use the same datasets as Tang et al.~\cite{tang2025llm4szz}, including DS\_LINUX, DS\_GITHUB, and DS\_APACHE.
DS\_LINUX contains 1,500 bug-fixing commits from the Linux kernel, sampled from the dataset of Lyu et al.~\cite{lyu2024evaluating}. 
DS\_GITHUB contains 361 bug-fixing commits from 285 GitHub repositories covering C and Java projects~\cite{rosa2021evaluating}. 
During our evaluation (March 2026), we removed 6 unavailable repositories, leaving 355 bug-fixing commits from 279 repositories.
DS\_APACHE contains 241 bug-fixing commits from five Apache projects~\cite{wen2019exploring}. 
In total, we evaluate \tool{} on 2,095 bug-fixing commits with 2,271 bug-inducing commits.






\subsection{Baselines}
\label{subsec:baselines}

We select nine SZZ algorithms discussed in Section~\ref{sec:background} as baselines to compare with \tool{}: B-SZZ~\cite{jacek2005when}, AG-SZZ~\cite{kim2006automatic}, MA-SZZ~\cite{da2016framework}, R-SZZ~\cite{davies2014comparing}, L-SZZ~\cite{davies2014comparing}, RA-SZZ~\cite{neto2018impact}, Neural-SZZ~\cite{tang2023neural}, TC-SZZ~\cite{lyu2024evaluating}, and LLM4SZZ~\cite{tang2025llm4szz}, following prior studies~\cite{tang2025llm4szz,lyu2024evaluating}. 
Neural-SZZ and RA-SZZ are limited to Java projects, so we just compare in the datasets with Java projects.
TG-SZZ~\cite{shi2026beyond} are excluded as they did not release their implementation.

To ensure a fair comparison, we run LLM4SZZ using the same LLM as \tool{} (GPT-5-mini~\cite{openai2025gpt5mini}), denoted as LLM4SZZ (5-mini) in our evaluation. 
We also include mini-swe-agent~\cite{yang2024swe} as a baseline, as it represents an agent-based software engineering framework that supports multi-step reasoning and tool use, making it suitable for complex repository analysis tasks. 
Overall, we compare \tool{} against eleven baselines.

\subsection{Evaluation Metrics}
\label{subsec:metrics}

Following prior work on SZZ evaluation~\cite{tang2025llm4szz}, 
we adopt precision, recall, and F1-score to measure bug-inducing commit identification accuracy. 
Let $\mathit{GT}(c_i)$ and $\mathit{Pred}(c_i)$ denote the ground-truth and predicted bug-inducing commit sets for bug-fixing commit $c_i$. 
We compute metrics over all commits as follows:

\begin{equation}
\mathit{Precision} = 
\frac{\sum_{i} |\mathit{GT}(c_i) \cap \mathit{Pred}(c_i)|}
{\sum_{i} |\mathit{Pred}(c_i)|}
\end{equation}

\begin{equation}
\mathit{Recall} = 
\frac{\sum_{i} |\mathit{GT}(c_i) \cap \mathit{Pred}(c_i)|}
{\sum_{i} |\mathit{GT}(c_i)|}
\end{equation}

\begin{equation}
F1 = 2 \cdot \frac{\mathit{Precision} \times \mathit{Recall}}
{\mathit{Precision} + \mathit{Recall}}
\end{equation}

\subsection{Implementation}
\label{subsec:implementation}

Using state-of-the-art code LLMs (e.g., Claude Opus 4.6~\cite{anthropic2026claudeopus46}) is impractical for large-scale evaluation due to their high cost. 
A single evaluation run on our three datasets would cost approximately \$400, making such models unsuitable for our study. 
While prior work~\cite{tang2025llm4szz} adopts GPT-4o-mini~\cite{openai2024gpt4omini}, we instead use GPT-5-mini~\cite{openai2025gpt5mini}, which provides stronger support for tool use and reasoning while remaining cost-effective.

All experiments are conducted on a server equipped with an AMD EPYC 7763 64-core processor and 252\,GB of RAM. 
We access LLM APIs via OpenRouter.\footnote{\url{https://openrouter.ai}}
We do not set the temperature to 0, as GPT-5 series models do not support temperature control~\cite{ashwinthandu2025gpt5temperature}. 
The agent is limited to a maximum of 15 interaction turns.

For the mini-SWE-agent (v2.2.5) baseline~\cite{yang2024swe}, we repurpose its bash-based repository interaction environment for BIC tracing instead of its default software repair workflow. 
The agent uses GPT-5-mini via OpenRouter and is run with a 300-second timeout per case. 
To ensure a fair comparison, it is provided with the same fix-side evidence as \tool{}: the repository path, the fix commit hash, and a filtered fix context derived from the fixing commit, with metadata trailers (e.g., \texttt{Fixes:}) removed. 
The agent is prompted to trace the origin of the bug using read-only git commands and to output the final BIC in a structured format.

\section{Results}
\label{sec:results}

\subsection{RQ1: Effectiveness of \tool{}}

Table~\ref{tab:overall} presents the results across three datasets. 
Following prior work~\cite{tang2025llm4szz}, we further split the GitHub dataset into C and Java subsets to analyze language-specific performance. 
This results in four datasets: DS\_LINUX (Linux), DS\_GITHUB-c (GitHub-C), DS\_GITHUB-j (GitHub-J), and DS\_APACHE (Apache).
To ensure a fair comparison, we run \tool{}, LLM4SZZ, and LLM4SZZ (5-mini) three times and report the average results. 
Due to its high cost and low performance, mini-swe-agent is evaluated with a single run on each dataset.


\begin{table*}[t]
\centering
\caption{
Performance comparison of methods for identifying bug-inducing commits on C and Java datasets. 
RA-SZZ and Neural-SZZ are only applicable to Java, so their results are marked as “--” for C datasets. 
Ablation results are shown in the bottom rows; 
For the main methods, the best value in each column is bold and the second-best is underlined.
}
\label{tab:overall}
\setlength{\tabcolsep}{5.5pt}
\renewcommand{\arraystretch}{1.12}
\small
\begin{tabular}{lccc ccc ccc ccc}
\toprule
\textbf{Method} &
\multicolumn{3}{c}{\textbf{DS\_LINUX}} &
\multicolumn{3}{c}{\textbf{DS\_GITHUB-c}} &
\multicolumn{3}{c}{\textbf{DS\_GITHUB-j}} &
\multicolumn{3}{c}{\textbf{DS\_APACHE}} \\
\cmidrule(lr){2-4}\cmidrule(lr){5-7}\cmidrule(lr){8-10}\cmidrule(lr){11-13}
& Prec. & Rec. & F1
& Prec. & Rec. & F1
& Prec. & Rec. & F1
& Prec. & Rec. & F1 \\
\midrule
B-SZZ     & 0.452 & 0.578 & 0.507 & 0.361 & \underline{0.656} & 0.466 & 0.285 & \underline{0.680} & 0.401 & 0.251 & \underline{0.435} & 0.318 \\
AG-SZZ    & 0.448 & 0.553 & 0.495 & 0.410 & 0.592 & 0.484 & 0.421 & 0.533 & 0.470 & 0.328 & 0.310 & 0.318 \\
MA-SZZ    & 0.421 & 0.538 & 0.472 & 0.335 & 0.624 & 0.436 & 0.239 & 0.560 & 0.335 & 0.307 & 0.345 & 0.329 \\
R-SZZ     & 0.583 & 0.448 & 0.507 & 0.671 & 0.582 & 0.620 & 0.538 & 0.467 & 0.500 & 0.497 & 0.288 & 0.364 \\
L-SZZ     & 0.560 & 0.430 & 0.486 & 0.486 & 0.422 & 0.452 & 0.492 & 0.427 & 0.457 & 0.366 & 0.211 & 0.267 \\
RA-SZZ    & --    & --    & --    & --    & --    & --    & 0.337 & 0.440 & 0.382 & 0.264 & 0.325 & 0.293 \\
Neural-SZZ& --    & --    & --    & --    & --    & --    & 0.556 & 0.486 & 0.520 & 0.563 & 0.364 & 0.442 \\
LLM4SZZ   & \underline{0.628} & 0.552 & 0.588 & 0.687 & 0.641 & 0.663 & 0.607 & 0.569 & 0.587 & 0.610 & 0.398 & 0.482 \\
\midrule
LLM4SZZ (5-mini) & 0.580 & \underline{0.607} & \underline{0.593} & \underline{0.691} & 0.655 & \underline{0.673} & \underline{0.696} & 0.600 & \underline{0.616} & \underline{0.623} & 0.401 & \underline{0.488} \\
mini-swe-agent & 0.865  & 0.010 & 0.149 & 0.870 & 0.139 & 0.240 & 1.000 & 0.160 & 0.276 & 0.633 & 0.054 & 0.099  \\
\midrule
\textsc{AgentSZZ} & \textbf{0.788} & \textbf{0.711} & \textbf{0.748} & \textbf{0.830} & \textbf{0.800} & \textbf{0.815} 
& \textbf{0.750} & \textbf{0.720} & \textbf{0.735} 
& \textbf{0.724} & \textbf{0.452} & \textbf{0.556} \\
\midrule
\quad w/o Tools       & 0.212 & 0.125 & 0.157 & 0.009 & 0.007 & 0.008 & 0.029 & 0.027 & 0.028 & 0.000 & 0.000 & 0.000 \\
\quad w/o Domain      & 0.798 & 0.641 & 0.711 & 0.889 & 0.775 & 0.828 & 0.754 & 0.653 & 0.700 & 0.739 & 0.393 & 0.513 \\
\quad w/o Compression & 0.792 & 0.746 & 0.768 & 0.812 & 0.789 & 0.801 & 0.841 & 0.773 & 0.806 & 0.706 & 0.469 & 0.564 \\

\bottomrule
\end{tabular}
\end{table*}

\tool{} achieves the highest recall and F1-score across all four datasets among eleven baselines. 
Compared to the prior state-of-the-art LLM4SZZ~\cite{tang2025llm4szz}, it improves F1 by 27.2\%, 22.9\%, 25.2\%, and 15.4\% on Linux, GitHub-C, GitHub-J, and Apache, respectively. 
It also achieves strong precision (0.788, 0.830, 0.750, and 0.724), corresponding to relative improvements of 25.5\%, 20.8\%, 23.6\%, and 18.7\% over LLM4SZZ. 
Although \tool{} has slightly lower precision, the extremely low recall of mini-swe-agent leads to misleadingly high precision due to its limited number of correct predictions.

Using the same backbone model (LLM4SZZ with GPT-5-mini), \tool{} improves F1 by 26.1\%, 21.1\%, 19.3\%, and 13.9\%, indicating that the gains stem from the agent design and domain knowledge rather than the choice of LLM. 
In contrast, mini-swe-agent~\cite{yang2024swe} performs poorly across all datasets (F1: 0.123, 0.123, 0.123, 0.011), largely due to unfocused exploration. 
While it retrieves many related commits using general-purpose tools (e.g., \texttt{git show}, \texttt{git log}, \texttt{git grep}), it rarely forms a stable tracing process toward the true BIC. 
Successful cases involve short tool chains and early convergence, whereas failures exhibit long exploratory sequences with repeated inspection but limited progress. 
Occasional command construction errors further degrade performance.

We compare our results with both LLM4SZZ through static analysis, as all run three times and the results show that \tool{} consistently outperforms both LLM4SZZ with significant and effective size.p < 0.001

\tool{} also substantially improves recall. 
Compared to B-SZZ, recall increases by 23.0\%, 22.0\%, 5.9\%, and 3.9\% on Linux, GitHub-C, GitHub-J, and Apache, respectively, while maintaining high precision. 
The smaller gain on Apache is due to its higher average number of bug-inducing commits per fix (1.46 vs.\ 1.04 for Linux), where \tool{} tends to identify a single bug-inducing commit, limiting recall in multi-bug-inducing commits cases.

\tool{} further demonstrates strong consistency and generalizability. 
Despite being designed based on Linux, it achieves substantial improvements on GitHub and Apache, and generalizes across both C and Java without dataset-specific tuning.

\begin{tcolorbox}[tile, size=fbox, boxsep=2mm, boxrule=0pt, top=0pt, bottom=0pt,
  borderline west={1mm}{0pt}{blue!50!white}, colback=blue!5!white]
  \textbf{Answers to RQ1}: 
 \tool{} outperforms state-of-the-art methods in precision, recall, and F1-score across all datasets, achieving F1 improvements of 14.5\%-27.2\%.
  It consistently outperforms LLM4SZZ under the same backbone, while mini-swe-agent performs poorly due to unfocused exploration.
\tool{} also generalizes well across datasets and programming languages.
\end{tcolorbox}

\subsection{RQ2. Efficiency of \tool{}}
\label{subsec:rq2}

Table~\ref{tab:cost} compares the efficiency of four LLM-based approaches in terms of runtime, token consumption, interaction turns, and monetary cost. 
All API calls are made via OpenRouter to ensure a fair comparison, and all values are averaged over three datasets.

\begin{table}[t]
\centering
\caption{Cost analysis of \tool{} in terms of tokens and time consumption.}
\label{tab:cost}
\setlength{\tabcolsep}{6pt}
\renewcommand{\arraystretch}{1.1}
\small
\begin{tabular}{lccccccc}
\toprule
\textbf{Method} & Time (s)  & Tokens & Turns & Cost (\$) \\
\midrule
LLM4SZZ & 26.2 & 13,033 & 11.2 & 0.003 \\
LLM4SZZ (5-mini) & 125.2 & 18,230 & 12.6 & 0.015 \\
mini-swe-agent & 183.3 & 330,327 & 18.6 & 0.035 \\
\tool{} & 39.1 & 34,657 & 5.7 & 0.009 \\
\midrule
\quad w/o Compression & 29.8 & 52,237 & 4.7 & 0.014 \\
\bottomrule
\end{tabular}
\end{table}

\tool{} achieves a strong balance between efficiency and effectiveness. 
Compared to LLM4SZZ, it incurs a modest increase in runtime (39.1s vs.\ 26.2s) and token usage (34,657 vs.\ 17,690), but requires substantially fewer interaction turns (5.7 vs.\ 11.2), resulting in a much higher per-turn information density (6,080 vs.\ 1,580 tokens/turn). 
This indicates that \tool{} extracts significantly more information per step through structured tool use, and converges with fewer, more targeted interactions.

When LLM4SZZ is powered by GPT-5-mini, it becomes substantially slower (160s), due to higher latency per API call, while its token usage and interaction patterns remain largely unchanged. 
This suggests that stronger model reasoning alone does not improve efficiency in static pipelines. 
In contrast, \tool{} achieves both faster execution and higher accuracy by leveraging structured tool interaction rather than relying solely on internal reasoning.
mini-swe-agent is significantly less efficient across all dimensions, consuming 4.7$\times$ more time, 9.5$\times$ more tokens, and 3.9$\times$ higher cost than \tool{}. 
Its higher turn count and excessive token usage reflect prolonged, unfocused exploration, a known limitation of open-ended bash-based agents. 
By contrast, \tool{} constrains the action space to curated git-based tools, enabling more directed and efficient investigation.

\tool{} achieves a strong balance between efficiency and performance. 
Compared to LLM4SZZ, it incurs only a modest increase in runtime (39.1s vs. 26.2s) and token usage, while using fewer interaction turns (5.7 vs. 11.2), indicating more efficient reasoning and faster convergence. 
Despite using more tokens than LLM4SZZ, the reduced number of turns suggests that \tool{} extracts more useful information per step through targeted tool usage. 
In contrast, mini-swe-agent is significantly less efficient, with much higher runtime (183.3s), token consumption (330k), and cost, due to prolonged and unfocused exploration. 
\tool{} reduces runtime by over 4$\times$ and token usage by nearly 10$\times$ compared to mini-swe-agent, while achieving substantially better accuracy. 


\begin{tcolorbox}[tile, size=fbox, boxsep=2mm, boxrule=0pt, top=0pt, bottom=0pt,
  borderline west={1mm}{0pt}{blue!50!white}, colback=blue!5!white]
  \textbf{Answers to RQ2}: 
  \tool{} achieves a strong trade-off between efficiency and effectiveness. 
  It requires only 5.7 turns on average, with a per-turn information density 3.8$\times$ higher than LLM4SZZ. 
  Compared to mini-swe-agent, it reduces runtime, token usage, and cost by 4.7$\times$, 9.5$\times$, and 3.9$\times$, respectively. 
\end{tcolorbox}

\subsection{RQ3. Key Components of \tool{}}
\label{subsec:rq3}

As shown in Table~\ref{tab:overall}, the last three rows present the results of our ablation study, where we evaluate the contribution of each key component of \tool{} by removing it individually. 
We consider three variants: 
(1)~\textit{w/o Tools}, which removes all tool access and forces the agent to rely solely on fix commit information without interactive investigation; 
(2)~\textit{w/o Domain}, which replaces the domain knowledge prompt with minimal task instructions; 
and (3)~\textit{w/o Compression}, which disables all context compression module.

\textbf{Structured tools are indispensable.}
Removing task-specific tools causes performance to collapse across all datasets, with F1 dropping from 0.748 to 0.157 on Linux, from 0.815 to 0.008 on GitHub-C, and to 0.000 on Apache. 
Without structured tool interfaces, the agent must compose commands, interpret raw outputs, and manage context within a limited turn budget—tasks that even strong LLMs struggle with. 
This confirms that structured tools are the most critical component of \tool{}.

\textbf{Domain knowledge consistently improves recall.}
Removing domain knowledge leads to consistent recall drops across all datasets, with the largest decrease observed on Apache ($-$13.0\%), where commits are larger and more complex. 
On GitHub-C, removing domain knowledge slightly increases precision (0.830$\to$0.889) but reduces recall, resulting in a marginal F1 increase (0.815$\to$0.828), while all other cases show an F1 decrease. 
Overall, domain knowledge promotes broader exploration—occasionally introducing false positives but consistently improving recall—while generalizing across both C and Java.

\textbf{Context compression improves efficiency with minimal impact on effectiveness.}
Disabling compression has limited impact on effectiveness, with F1 changes within $\pm$2\% on the three larger datasets. 
In contrast, its impact on efficiency is substantial: token consumption decreases by 33.6\% (52,237$\to$34,657) and cost by 35.7\% (\$0.014$\to$\$0.009). 
This highlights the importance of compression for practical deployment. 
We focus on the three larger datasets due to higher variance in GitHub-J.

\begin{figure*}[t]
  \centering
  \includegraphics[width=0.9\linewidth]{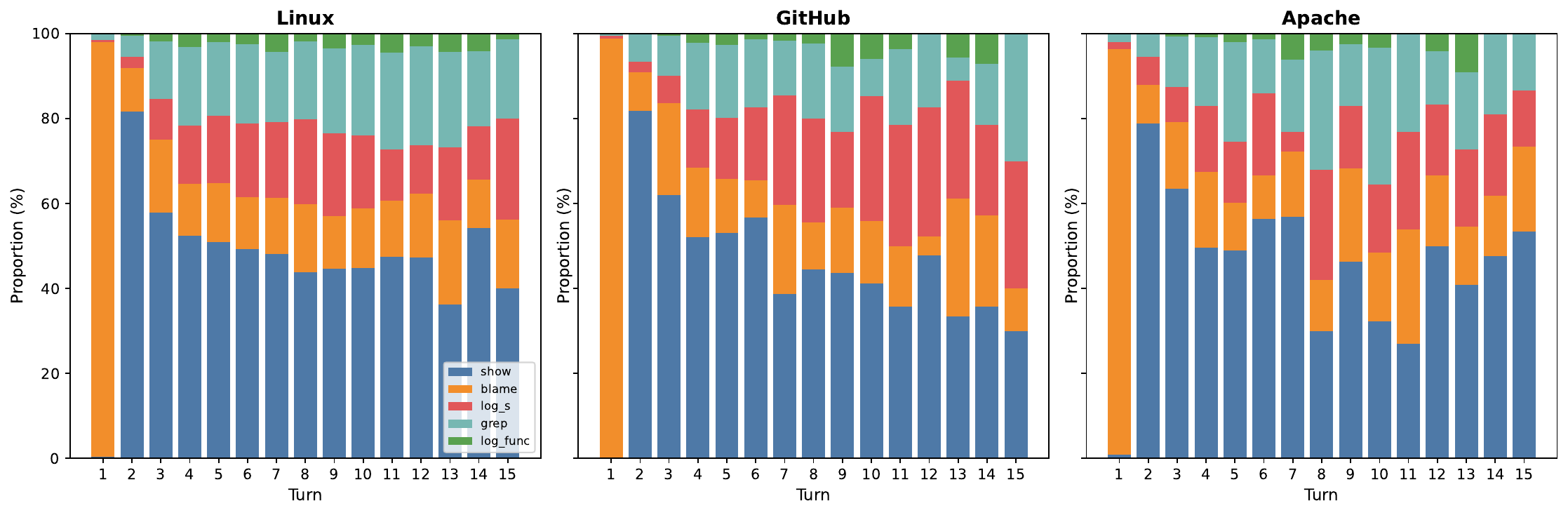}
  \caption{Distribution of tool usage across investigation steps in \tool{}. 
  The values are normalized proportions per turn.}
  \label{fig:tool_usage}
\end{figure*}

Figure~\ref{fig:tool_usage} shows the distribution of tool usage across investigation turns. 
Most cases are resolved within 4--6 turns, with \texttt{git\_show} and \texttt{git\_blame} accounting for approximately 40\% and 18\% of tool calls, respectively. 
Early turns are dominated by \texttt{git\_blame}, while \texttt{git\_show} peaks during candidate inspection. 
In later turns, the agent increasingly relies on broader search tools (\texttt{git\_log\_s} and \texttt{git\_grep}), reflecting a transition from local blame to global search when initial evidence is insufficient. 
\texttt{git\_log\_func} is used sparingly as a fallback for function-level tracing.
Exploratory tools (\texttt{git\_grep}, \texttt{git\_log\_s}) appear more frequently in harder scenarios (e.g., omission, refactor-penetration, cross-file cases), where they serve as escalation mechanisms for cross-file and semantic tracing. 
\texttt{git\_log\allowbreak\_func} plays a minor role as a specialized fallback.

\begin{tcolorbox}[tile, size=fbox, boxsep=2mm, boxrule=0pt, top=0pt, bottom=0pt,
  borderline west={1mm}{0pt}{blue!50!white}, colback=blue!5!white]
  \textbf{Answers to RQ3}: Designed Tools are indispensable, with F1 dropping to near zero without them. 
  Domain knowledge improves recall, while context compression reduces token consumption by over 30\% with minimal impact on effectiveness.
\end{tcolorbox}

\subsection{RQ4. Challenging Scenarios}
\label{subsec:rq4}

To evaluate \tool{} in challenging scenarios, we focus on bug-fixing commits categorized as cross-file and ghost cases (Table~\ref{tab:hard_case_recall}). 
Following Lyu et al.~\cite{lyu2024evaluating}, a case is defined as cross-file if none of its bug-inducing commits appear in the history of files modified by the fix (via \texttt{git log}), and as a ghost case if the bug-fixing commit contains no deleted or modified lines.

\begin{table*}[t]
\centering
\caption{Distribution, recall, and relative improvement of \tool{} over LLM4SZZ on cross-file and ghost commit cases.}
\label{tab:hard_case_recall}
\small
\setlength{\tabcolsep}{4.5pt}
\begin{tabular}{lrrr ccc ccc}
\toprule
\multirow{2}{*}{\textbf{Dataset}} & \multirow{2}{*}{\textbf{Total}} & \multirow{2}{*}{\textbf{Cross-file}} & \multirow{2}{*}{\textbf{Ghost}} & \multicolumn{3}{c}{\textbf{Cross-file Recall}} & \multicolumn{3}{c}{\textbf{Ghost Recall}} \\
\cmidrule(lr){5-7} \cmidrule(lr){8-10}
 &  &  &  & \tool{} & LLM4SZZ & Improv. & \tool{} & LLM4SZZ & Improv. \\
\midrule
Apache  & 243  & 34 (14.0\%)  & 10 (4.1\%)   & 18.60 &  9.30 & +100.0\% & 53.33 & 40.00 & +33.3\% \\
GitHub  & 355  & 33 (9.3\%)   & 37 (10.2\%)  & 24.24 &  6.06 & +300.0\% & 70.27 & 48.65 & +44.4\% \\
Linux   & 1500 & 209 (13.9\%) & 263 (17.5\%) & 42.72 & 16.43 & +160.1\% & 64.18 & 39.93 & +60.7\% \\
\midrule
Overall & 2104 & 276 (13.1\%) & 310 (14.7\%) & 37.02 & 14.19 & +160.9\% & 64.38 & 40.94 & +57.2\% \\
\bottomrule
\end{tabular}
\end{table*}

The results show that our approach significantly outperforms LLM4SZZ in both cross-file and ghost commit scenarios. 
For cross-file cases, \tool{} achieves recall of 18.60\%, 24.24\%, and 42.72\% on Apache, GitHub, and Linux, respectively, corresponding to improvements of 100.0\%, 300.0\%, and 160.1\%.

LLM4SZZ can handle a small number of cross-file cases, but its success mainly stems from indirect coverage rather than explicit cross-file reasoning. 
Its pipeline remains centered on a single fix-side file, where buggy statements are extracted and traced within the same file history. 
As a result, it only succeeds when sufficient local signals remain (e.g., due to file renaming or residual context in the patch). 
In contrast, \tool{} is not restricted to a single file history. 
Starting from a \texttt{git\_blame} $\rightarrow$ \texttt{git\_show} backbone, it can escalate to \texttt{git\_grep} and \texttt{git\_log\_s} to bridge symbol-level, statement-level, and cross-file relationships, enabling reliable transitions from the fix location to the true origin.

A similar pattern explains the gap on ghost commits. 
LLM4SZZ relies on explicit line-level signals from the fixing patch to recover candidate commits. 
When no lines are deleted or modified, this signal becomes unavailable, significantly weakening the pipeline. 
By contrast, \tool{} leverages broader context, semantic reasoning, and tool-assisted exploration to identify the bug-inducing commit even without direct line-level evidence, making it substantially more robust in ghost scenarios.

\begin{tcolorbox}[tile, size=fbox, boxsep=2mm, boxrule=0pt, top=0pt, bottom=0pt,
  borderline west={1mm}{0pt}{blue!50!white}, colback=blue!5!white]
  \textbf{Answers to RQ4}: Our approach significantly outperforms LLM4SZZ in challenging scenarios, improving recall by up to 300\% on cross-file cases and 60\% on ghost commits. 
\end{tcolorbox}

\section{Discussion}
\label{sec:discussion}

\subsection{Failure Analysis}
\label{subsec:failure}

We conduct a qualitative error analysis on a sample of 50 incorrect cases from Apache, GitHub, and Linux. 
The dominant failure mode is misattribution among plausible related commits. 
Specifically, 24/50 errors are local semantic misattributions, where the agent identifies a nearby plausible commit but fails to trace back to the earliest bug-inducing commit, and 11/50 stop at later related commits (e.g., propagation, follow-up, or refactoring steps). 

Cross-file localization misses account for 5/50 cases, indicating limited pivoting beyond the fix-side file history in some instances. 
Empty-output failures are less frequent (8/50) but involve substantially longer trajectories, suggesting continued exploration without successful convergence. 
Notably, 40/50 errors are assigned high confidence, indicating that the remaining challenge lies in precise historical attribution rather than repository exploration.

These patterns are illustrated by representative cases. 
In an Apache case,\footnote{\url{https://github.com/apache/accumulo/commit/a2c2d38aa}} the agent remained anchored to the fix-side file history and selected a plausible blame-linked commit instead of the true cross-file origin. 
In a GitHub case\footnote,{\url{https://github.com/sebastiaanschool/sebastiaanschool-Android/commit/2454879bfb}} the agent chose a highly relevant but later related commit rather than the earliest introducing one. 
In a Linux ghost commit case,\footnote{\url{https://github.com/torvalds/linux/commit/7ecb37f62fe5}} weak fix-side signals led to a high-confidence but non-earliest attribution.

\subsection{Threats to Validity}
\label{subsec:threats}

\textbf{Internal Validity.}
LLMs may produce non-deterministic outputs even with temperature set to 0, potentially affecting reproducibility. 
We mitigate this by running each experiment three times and reporting average results. 
The domain knowledge in our prompt was derived from 50 Linux kernel cases. 
Although results show strong cross-dataset and cross-language generalizability, the prompt may encode Linux-specific patterns that are less relevant to other projects. 
Future work could explore project-adaptive prompt generation to further improve robustness. 
The design process involved two authors interacting with an LLM to simulate agent behavior, introducing potential subjectivity in tool and prompt design. 
To mitigate this, the 50 design cases were excluded from all evaluation datasets, and all design choices were validated through ablation studies (RQ3). 
Prior work has shown that potential data leakage in similar settings does not materially affect evaluation outcomes. 
Therefore, we do not conduct additional experiments on this aspect, as it is orthogonal to the focus of this study. 

In addition, the performance of \tool{} depends on the quality of tool outputs and repository metadata (e.g., commit history and file structure). 
Incomplete or inconsistent repository information may affect the accuracy of tracing results. 
\tool{} attempts prefix matching at progressively shorter lengths (12, 10, 8, 7 characters) to recover from minor errors. 

\textbf{External Validity.}
Our evaluation spans three datasets, two programming languages (C and Java), and over 285 repositories, covering a broader scope than most prior SZZ studies. 
However, generalizability to other languages (e.g., Python, JavaScript) and project types (e.g., web or mobile applications) remains to be validated. 
We use developer-annotated ground truth, the highest-quality labeling available for SZZ evaluation. 
However, annotations may still contain errors or ambiguities, especially for complex bugs involving multiple commits. 
Additionally, DS\_GITHUB excludes six repositories due to unavailability during our evaluation period, which may slightly affect comparability with prior work. 
Finally, our evaluation focuses on open-source projects; the applicability of \tool{} to proprietary or industrial codebases with different development practices remains an open question.


\section{Conclusion and Future Work}
\label{sec:conclusion}

In this paper, we propose \textsc{AgentSZZ}, an agent-based framework that leverages LLM-driven agents to identify bug-inducing commits through interactive repository investigation. 
Unlike static \textit{git blame} pipelines, \textsc{AgentSZZ} combines task-specific tools, domain knowledge, and a ReAct-style loop to enable adaptive, evidence-driven exploration of the repository. 
A structured compression module further reduces token consumption by over 30\% with negligible impact, improving efficiency without sacrificing effectiveness.

Extensive experiments on three developer-annotated datasets show that \textsc{AgentSZZ} outperforms all eleven baselines, achieving F1 improvements of 15.4\%--27.2\% over LLM4SZZ. 
The gains are most pronounced in challenging scenarios: recall improves by up to 300\% on cross-file cases and 60\% on ghost commits, where static tracing methods often fail. 
Ablation studies show that task-specific tools are indispensable, domain knowledge significantly improves recall, and the overall architecture generalizes well across multiple LLM backbones. 
These results suggest that structured tool interaction, rather than purely model-based reasoning, is key to effective and efficient bug-inducing commit identification.

Several directions remain for future work. 
The dominant failure mode is misattribution among plausible commits rather than localization errors, suggesting the need for finer-grained temporal reasoning or stronger causal verification. 
For example, incorporating execution-based validation could help distinguish commits that introduce faults from those that merely expose them.
\tool{} currently tends to identify a single BIC per fix, limiting recall in multi-BIC cases (e.g., DS\_APACHE). 
Future work could extend the framework to support multiple BICs by modeling dependencies across commits.
Future work could also explore automated domain knowledge extraction and integrate execution or static analysis to better verify causal relationships.


\balance{}

\bibliographystyle{ACM-Reference-Format}
\bibliography{reference}


\end{document}